\documentclass[twocolumn,preprintnumbers,floatfix,prb,superscriptaddress]{revtex4}
\usepackage{graphicx}
\usepackage{graphicx}
\usepackage{dcolumn} 
\usepackage{bm}
\usepackage{epsfig}
\begin{document} 

\title{Enhanced thermoelectric response of hole-doped La$_2$NiO$_{4+\delta}$ by ab initio calculations}

\author{Victor Pardo}
 \email{victor.pardo@usc.es}
\affiliation{
Departamento de F\'{\i}sica Aplicada, Universidade
de Santiago de Compostela, E-15782 Santiago de Compostela,
Spain
}
\affiliation{
Instituto de Investigaci\'{o}ns Tecnol\'{o}xicas, Universidade
de Santiago de Compostela, E-15782 Santiago de Compostela,
Spain
}

\author{Antia S. Botana}
\affiliation{
Departamento de F\'{\i}sica Aplicada, Universidade
de Santiago de Compostela, E-15782 Santiago de Compostela,
Spain
}
\affiliation{
Instituto de Investigaci\'{o}ns Tecnol\'{o}xicas, Universidade
de Santiago de Compostela, E-15782 Santiago de Compostela,
Spain
}

\author{Daniel Baldomir}
\affiliation{
Departamento de F\'{\i}sica Aplicada, Universidade
de Santiago de Compostela, E-15782 Santiago de Compostela,
Spain
}
\affiliation{
Instituto de Investigaci\'{o}ns Tecnol\'{o}xicas, Universidade
de Santiago de Compostela, E-15782 Santiago de Compostela,
Spain
}


\begin{abstract}

Thermoelectric properties of the system La$_2$NiO$_{4+\delta}$ have been studied ab initio. Large Seebeck coefficient values are predicted for the parent compound, and to some extent remain in the hole-doped metallic phase, accompanied of an increase in the conductivity. This system, due to its layered structure would be a suitable candidate for an improvement of its thermoelectric figure of merit by nanostructurization in thin films, that has already been shown to increase the electrical conductivity ($\sigma$). Our calculations show that in the region around La$_2$NiO$_{4.05}$ the system has a large thermopower at high temperatures and also a substantially increased $\sigma$. Films grown with this low-doping concentration will show an optimal relationship between thermopower and $\sigma$. This result is obtained for various exchange-correlation schemes (correlated, uncorrelated and parameter-free) that we use to analyze the electronic structure of the hole-doped compound.

\end{abstract}

\maketitle

\newpage

\section{Introduction}

In recent times, the studies on the thermoelectric properties of various compounds both using experiments, theoretical and computational tools has focused in the search for new functional materials that could be used in the fabrication of energy-recovery devices based on a thermoelectric effect. Oxides are promising in principle because of their chemical stability and possibility of usage in high temperature devices. Particularly, misfit layered cobaltates have drawn significant attention.\cite{review_nat_therm, naxcoo_singh, prb_ca3co4o9, prb_misfit, jssc} Large thermopower together with metallic conductivities place that family as one of the most relevant and thoroughly studied within the oxides. One electronic structure reason behind their interesting thermoelectric properties is the existence of two electronic systems within those cobaltates.\cite{prb_eg_a1g} Due to the trigonally distorted octahedral environment, the t$_{2g}$ bands are split between a highly mobile wide band that drives metallic conduction (of e$_g^*$ character) and a more localized band (with a$_{1g}$ parentage), narrow in nature that provides a large Seebeck coefficient.\cite{naxcoo2_terasaki}

The search for new oxide compounds that could present a similar thermoelectric response is an interesting issue to be pursued. In this paper we will argue that a possible candidate family for that sake are a particular kind of layered nickelates: hole-doped La$_2$NiO$_4$. This compound and others related to it have drawn the attention of the community for various reasons: i) La$_2$NiO$_4$ is isostructural with the parent compound of the high-T$_c$ cuprates:\cite{la2nio4_struct} NiO$_2$ planes with a square planar coordination around the metal cations form the building blocks of this layered material, with La acting as spacer between the layers, ii) an interesting temperature-induced metal-insulator transition (MIT) occurs in it,\cite{la2nio4_mit} iii) it is possible to lower further the valence of the Ni cation in a layered structure to compare it with superconducting cuprates, not only in terms of structure but also with the same electron count.\cite{lanio_2006,lanio_2007,lanio_curro_1,lanio_polt1,lanio_curro_2,la4ni3o8_vpardo,la3ni2o6_prl,la4ni3o8_goodenough} Back in the eighties, the MIT was explained by Goodenough and Ramasesha\cite{la2nio4_goodenough} as occurring in a system where two different $d$-states coexist around the Fermi level, having clearly distinct band widths.  This is reminiscent of the above mentioned cobaltates, but in this case two different e$_g$ bands exist, a more localized d$_{z^2}$ and a more itinerant $d_{x^2-y^2}$ band. These yield reasonable conductivities, and also due to the existence of a more localized set of narrow bands, interesting values of the thermopower could emerge.

Experimental studies of the Seebeck coefficient in La$_2$NiO$_4$ show promising values exceeding 100 $\mu$V/K at room temperature,\cite{la2nio4_Seebeck_delta} (the typical minimum value for a material to be considered a high performance thermoelectric) even though most reports with oxygen-rich samples show smaller values (we will argue below what the reason for this is). The electrical conductivity is larger in the NiO$_2$ planes,\cite{la2nio4_sigma_singlecrystals} which is the expected behavior for such a layered structure.  To obtain an estimate of the thermoelectric performance of a material, it is common to use the so-called figure of merit, which is defined as the dimensionless quantity $zT$=$\sigma$TS$^2$/$\kappa$. $zT$ $>$ 1 is required for applications.\cite{zt_minimum}
With the best values obtained experimentally for $S$ $\sim $100 $\mu$V/K,\cite{la2nio4_Seebeck_delta} $\rho$ $\sim$ 5 m$\Omega$.cm \cite{la2nio4_films_sigma,la2nio4_sigma_singlecrystals} and  $\kappa$ $\sim$ 7.5 W/m.K,\cite{la2nio4_kappa} a $zT$ at room temperature of about 0.01 can be estimated, which is comparable to other promising compounds such as CrN,\cite{camilo_apl,camilo_prb} particularly so because of the large room for improvement, some of which we will discuss in this paper. In particular, the layered structure of the compound would be prone to yield better conductivity when grown in the form of thin films, and this has been shown in the past.\cite{la2nio4_films_sigma} A thin film geometry would hamper thermal conductivity along the c-axis, which would eventually improve the thermoelectric figure of merit as long as the thermopower and the mobility are not reduced.\cite{bi2te3_1} Also, oxygen excess can lead to hole doping and a corresponding increase in electrical conductivity. A vast amount of experimental information exists on the thermoelectric and transport properties of La$_2$NiO$_4$. However, an exact account of the oxygen content is necessary to obtain good systematics, and this is not easily found in literature, probably due to complications in the correct determination and stabilization of a particular oxygen content. 

In this paper we will explore a possible path towards an improved thermoelectric response via oxygen doping. We will discuss in particular how moving the system into the more itinerant limit by hole doping can reduce further the importance of the phononic part of the thermal conductivity (that works against a good thermoelectric performance) relative to the electronic part, and also improving the figure of merit if a decrease in the thermopower is prevented. Concerning the magnetic order, it has been shown that La$_2$NiO$_4$ is an antiferromagnet with an in-plane ordering such that an antiferromagnetic (AF) interaction between nearest neighbor Ni atoms is stabilized. The commensurability of the ordering has been put into question,\cite{la2nio4_magn_short_range} but careful studies with respect to the oxygen content show how the N\'eel temperature is significant even at values of $\delta$ $\sim$ 0.1,\cite{la2nio4_TN_delta} and could vanish at $\delta$ $\sim$ 0.14.\cite{la2nio4_Seebeck_delta} We will focus our calculations in that doping level ($\delta$ $<$ 0.15), close to the localized electron limit where magnetic order is present at low temperature and our DFT calculations, that assume a magnetically ordered phase at zero temperature can describe the system more accurately. All calculations presented here will have that type of ordering imposed.

The paper is organized as follows: Section \ref{compdet} will present the calculation details, Section \ref{results} will describe the electronic structure, analyze the thermoelectric properties calculated for various oxygen contents, and finally we will summarize the main conclusions of the work in Section \ref{summary}.

\section{Computational Procedures}\label{compdet}

Our electronic structure calculations were  performed within density functional
theory\cite{dft,dft_2} using the all-electron, full potential code {\sc wien2k}\cite{wien}
based on the augmented plane wave plus local orbitals (APW+lo) basis set.\cite{sjo}
We will present calculations of the electronic structure and transport properties obtained with various exchange-correlation functionals and for several doping levels. For the exchange-correlation functional in the calculations at U= 0, we have used the Perdew-Burke-Ernzerhof version of the generalized gradient approximation\cite{gga} (GGA).

To deal with  strong correlation effects that are widely acknowledged to play an important role in this type of nickelates,
we apply the LDA+$U$ scheme \cite{sic1,sic2} that incorporates an on-site Coulomb repulsion U and Hund's rule coupling strength J$_H$ 
for the Ni $3d$ states. For the uncorrelated part of the exchange-correlation functional we used the local density 
approximation (LDA)\cite{lda} in that case.  The LDA+$U$ scheme utilized is the so-called ``fully localized limit"\cite{fll} (FLL).
A description of the results obtained at different values of U (in a reasonably broad range 4.5-8.5 eV for the Ni cations) was performed, but results presented here will be for U= 7 eV. No significant changes in the main conclusions of the paper are found within that range of U values. For a discussion of the effects of a different choice of U and LDA+$U$ method in similar layered nickelates, see Ref. \onlinecite{vpardo_sst}. The value chosen for the on-site Hund's rule strength is J = 0.68 eV and is kept fixed. 

Also, the recently developed Tran-Blaha modified Becke-Johnson (TB-mBJLDA) potential was used.\cite{TBmBJLDA} This potential has been shown to provide an accurate account of the electronic structure of correlated compounds\cite{crn_antia} using a parameter-free description, also yielding accurate band gaps for most semiconductors.

All calculations were fully converged 
with respect to all the parameters used. In particular, we used R$_{mt}$K$_{max}$= 7.0, a k-mesh of 10 $\times$ 10 $\times$ 4, 
and muffin-tin radii of 2.35 a.u. for La, 1.97 a.u. for Ni and 1.75 a.u. for O. 

The transport properties were calculated using a semiclassical solution based on the Boltzmann transport theory within the constant scattering time approximation by means of the BoltzTrap code, that uses the energy eigenvalues calculated by the {\sc wien2k} code.\cite{boltztrap} In this case, denser k-meshes are required, in our case up to 40 $\times$ 40 $\times$ 15 to reach convergence. The constant scattering time approximation assumes the relaxation time $\tau(\epsilon)$ as energy-independent. This results in  an expression of both the thermopower and the thermoelectric figure of merit with no dependence on $\tau$ (it can be directly obtained from the band structure without any assumed parameters). This approximation has been used succesfully to describe several thermoelectric materials.\cite{singh_1, singh_2, singh_3,singh_4,singh_5}

To simulate small doping values we have used the virtual crystal approximation (VCA).\cite{vca} This consists in modifying the total number of electrons in the system (also the total atomic number for charge neutrality) to simulate a particular doping level.

\section{Results}\label{results}

\subsection{Basic description of the electronic structure}

\begin{figure}[ht]
\begin{center}
\includegraphics[width=0.55\columnwidth,draft=false]{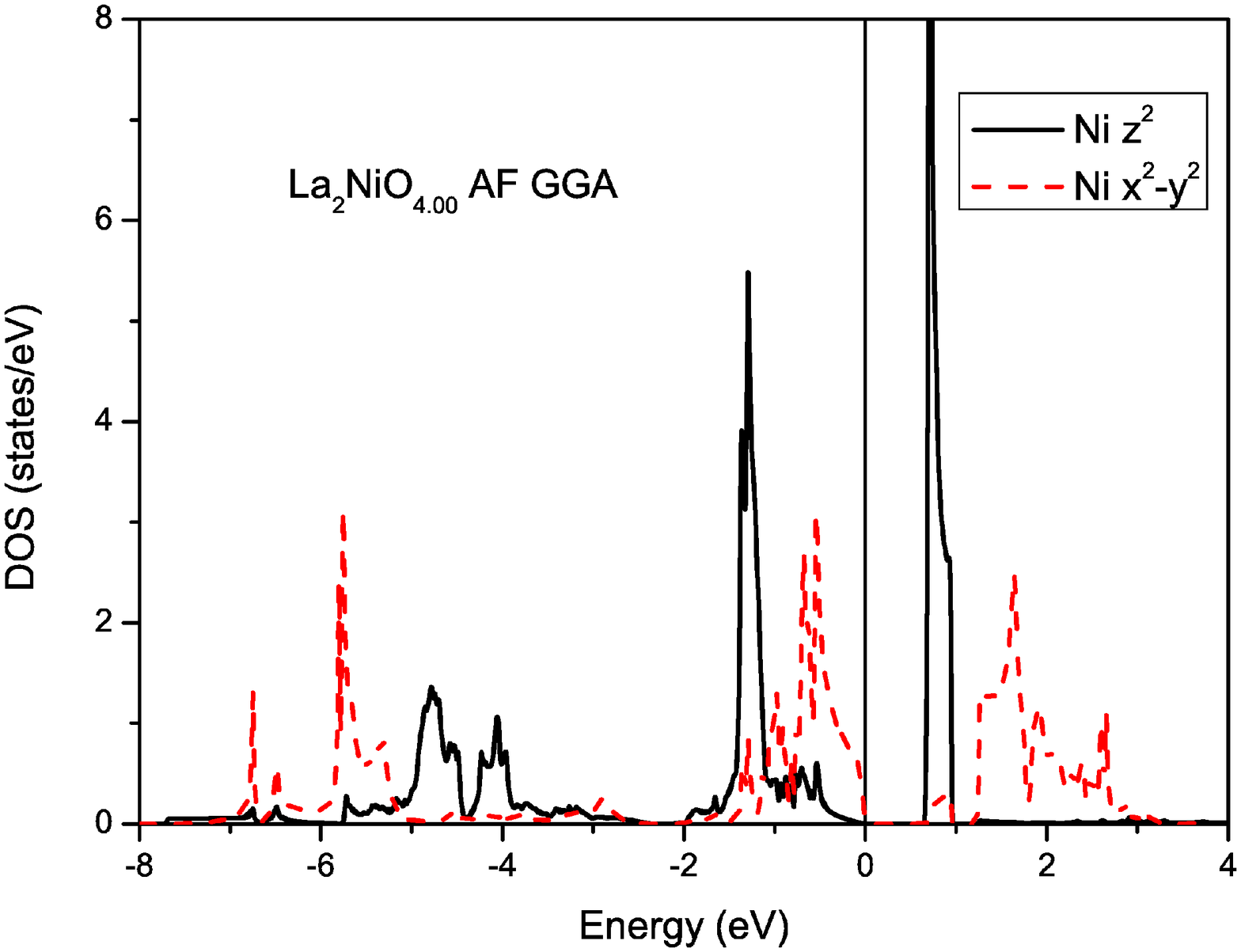}
\includegraphics[width=0.55\columnwidth,draft=false]{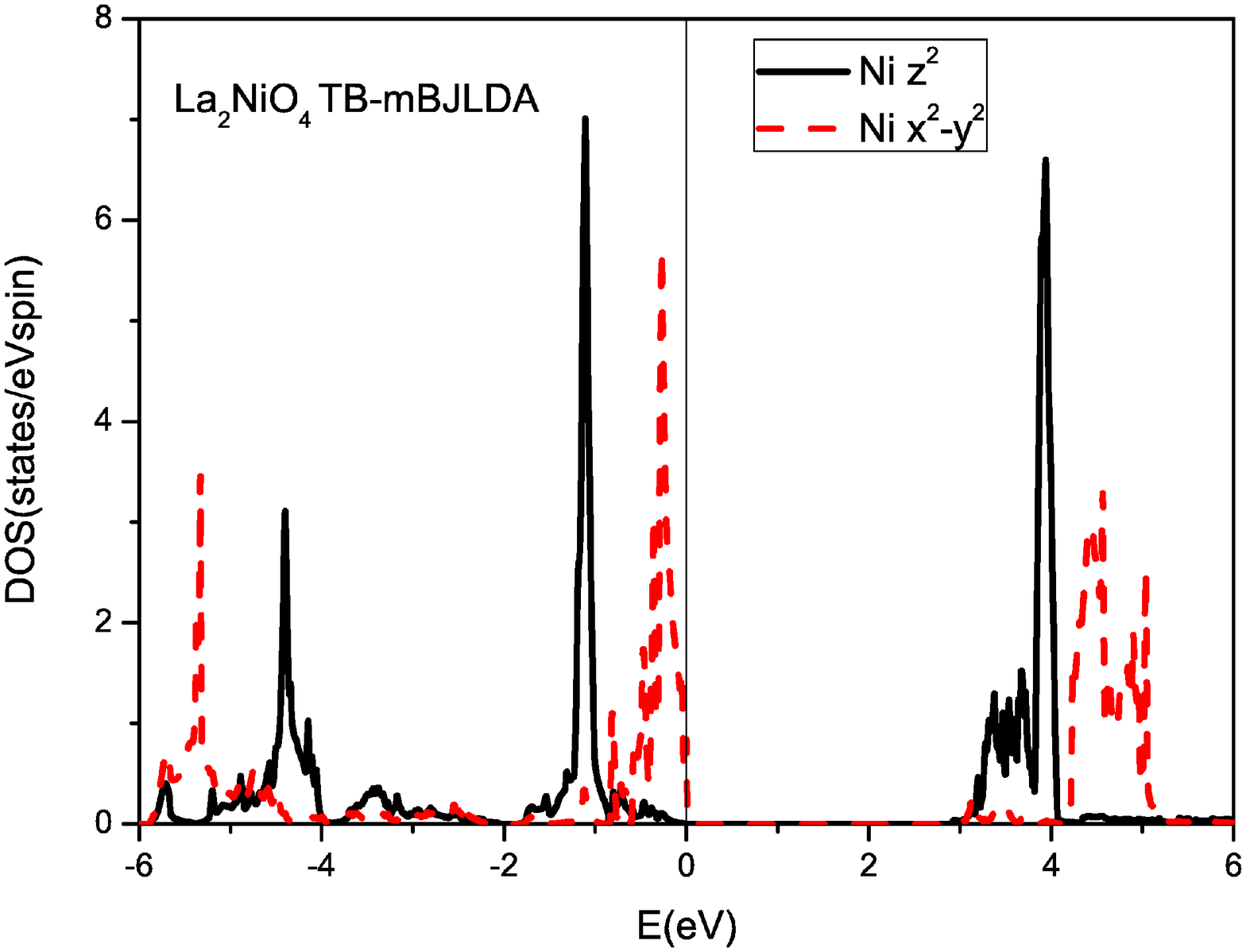}
\includegraphics[width=0.61\columnwidth,draft=false]{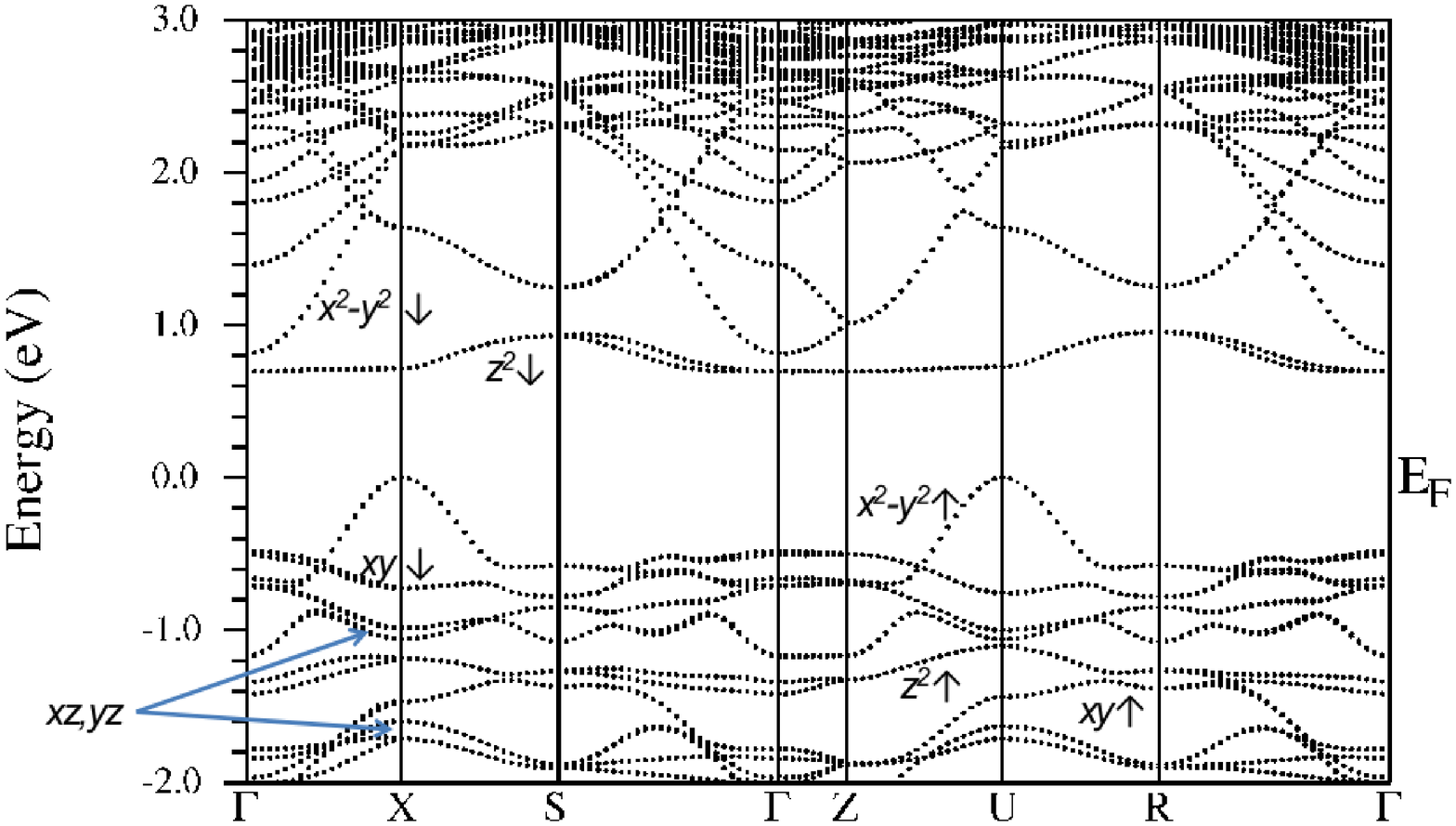}
\caption{(Color online.) DOS of the Ni e$_g$ levels in La$_2$NiO$_4$ ordered AF within GGA and TB-mBJLDA. Observe the more localized $d_{z^2}$ band and the broader $d_{x^2-y^2}$, particularly noticeable in the GGA calculation. Lower panel shows the band structure in the GGA calculation. The t$_{2g}$ bands are in the same energy region as the majority d$_{z^2}$ band.}\label{dos}
\end{center}
\end{figure}

A simple ionic picture of La$_2$NiO$_4$ gives a formal valence Ni$^{2+}$:d$^8$, which is always a S= 1 ion. Typically, for this type of low-valence nickelate, with the Ni$^{2+}$ cations in a square planar environment, one would expect the $xz/yz$ doublet fully occupied, and a higher position of the $xy$ singlet competing in occupation with the e$_g$ bands (which are also split and will have significantly different band widths). The typical situation would be a lower $xy$ band fully occupied and two e$_g^{\downarrow}$ unoccupied bands, being the band gap\cite{la2nio4_gap} of a distinct $d$-$d$ character.

Figure \ref{dos} shows the calculated density of states (DOS) for La$_2$NiO$_4$ in an AF solution, both for GGA and a TB-mBJLDA scheme together with the GGA band structure in the lower panel. Similar DOS curves are obtained with the various functionals, the size of the gap being the main difference, and also some differences exist in the particular bandwidths and position of the bands, but even calculations with U= 0 yield an insulating solution as long as in-plane AF order is set. In the plots, we only show the e$_g$ states of the Ni cations since these have the dominant contribution above and immediately below the Fermi level (and we want to see them with some degree of clarity). Furthermore, due to the square planar oxygen environment, the e$_g$ states are significantly split, and differ substantially not only in position but in width. The $d_{x^2-y^2}$ band is higher in energy due to crystal-field splitting caused by the absence of apical oxygens around the Ni cations, but it is slightly wider, whereas the $d_{z^2}$ band is lower but it is composed of more localized electrons (it is narrower). One can see this very clearly in the band structure figure. The $d_{z^2}$ bands appear both in the GGA and TB-mBJLDA calculations much narrower, with less than 1 eV band width, being the corresponding width of the $d_{x^2-y^2}$ band at least 2 eV. In the band structure figure (lower panel), we can also see  that not only the e$_g$ levels are in the vicinity of the Fermi level but also the t$_{2g}$ states are placed somewhat close to it, and are actually in the same enery region as the majority-spin d$_{z^2}$ band.

This is a narrow-gap oxide (of only 70 meV\cite{la2nio4_gap} according to resistivity measurements), but calculations show the gap is substantial (0.6 eV for GGA, 1 eV for LDA+$U$ at U= 7 eV and 2 eV for TB\-mBJLDA). Other narrow-gap oxides show this type of band gap overestimation coming out from calculations (CrN,\cite{crn_antia} Sr$_6$Co$_5$O$_{15}$\cite{sr6co5o15_antia}) and still the calculations of the transport properties agree well with experiments. One possibility for this could be the fact that the low gap value obtained from resistivity measurements comes from shallow states in the gap, being the actual band gap much larger, as happens in CrN.\cite{crn_antia} In the case of La$_2$NiO$_4$, a small band gap is consistent with the temperature-induced MIT observed. The MIT in this system has been described as a temperature-induced broadening of the $d_{x^2-y^2}$ band that eventually overlaps the occupied with the unoccupied band of opposite spin yielding metallic behavior at high temperature in the parent compound.\cite{la2nio4_mit} Our calculations confirm the description of the electronic structure that Goodenough and Ramasesha have done in the past,\cite{la2nio4_goodenough} although we see that the t$_{2g}$ bands are located together with the majority $d_{z^2}$ band in the vicinity of the Fermi level. The GGA DOS we show would anticipate that a sufficient temperature-induced broadening of the $x^2-y^2$ band would place the unoccupied band edge below that of the unoccupied $z^2$ band, which is significantly narrow. Something similar can be expected for the TB-mBJLDA calculation, but the largely overestimated gap makes the analysis a bit more adventurous.

\subsection{Tuning the electronic structure via hole-doping}

As mentioned before, if somehow one can complement metallic conductivity with the maintenance of a large thermopower and a reduced thermal conductivity, the material could then be comparable to other oxides in thermoelectric performance.\cite{naxcoo_singh} Let us see how this can be done in this particular case. As can be seen, the DOS rises rapidly right below the top of the valence band ($d_{x^2-y^2}$-character) which is generally favorable for the thermopower. A similar heavy-electron band appears somewhat also in the electron-doping region just above the bottom of the conduction band (in that case, of $d_{z^2}$ parentage) although with less dispersive character. Hence, the more localized $d_{z^2}$ band should contribute substantially to the Seebeck coefficient in the hole-doping region as well, but its DOS peak is very far from the Fermi level to be attained by a relatively small hole-doping level. This would instead move the chemical potential into the occupied $x^2-y^2$ band with a noticeable increase in electrical conductivity. This band could also be sufficiently narrow to retain a substantial value of the thermopower if the chemical potential lies in a region inside the valence band comprising a large derivative of the density of states.

\begin{figure*}[ht]
\begin{center}
\includegraphics[width=0.66\columnwidth,draft=false]{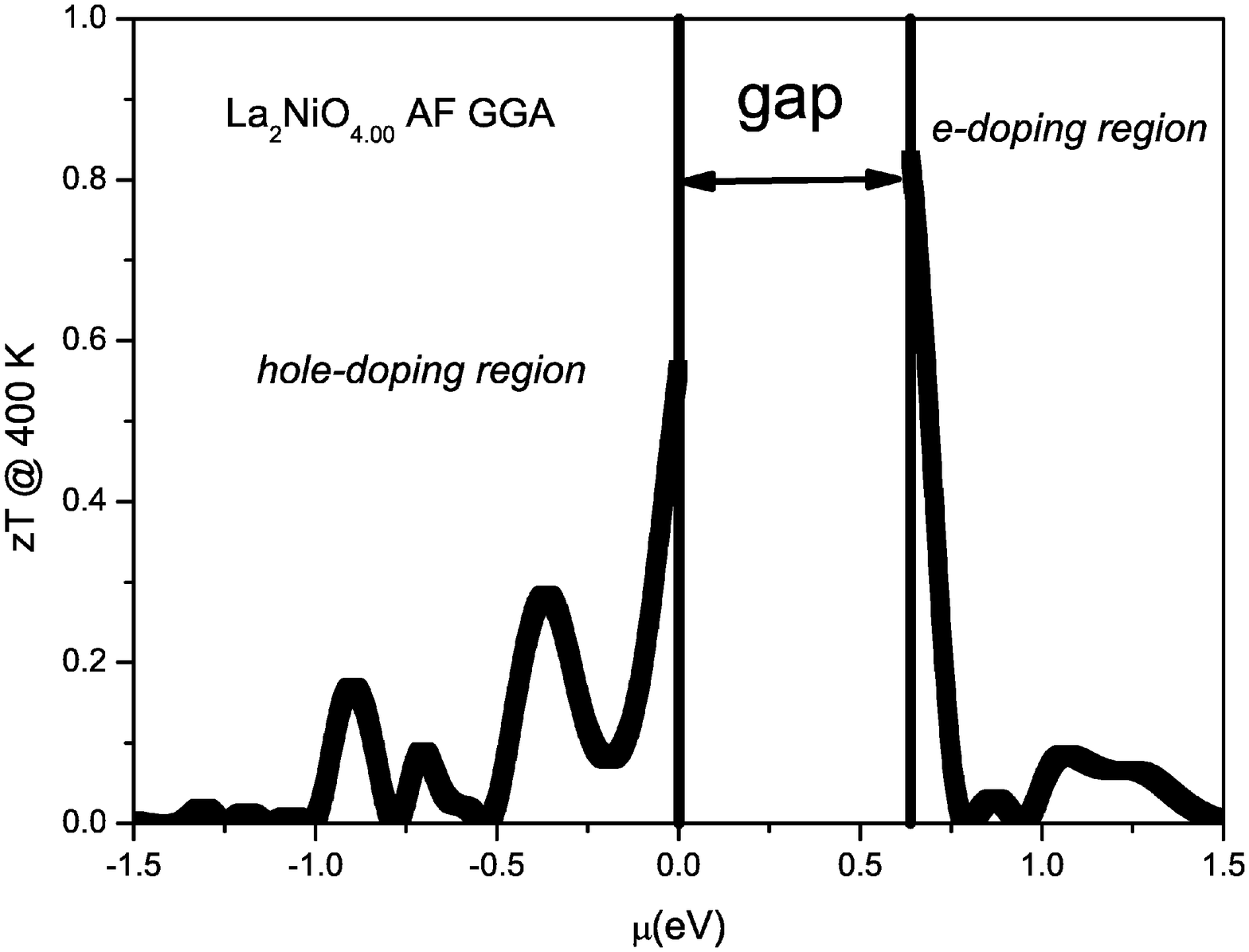}
\includegraphics[width=0.66\columnwidth,draft=false]{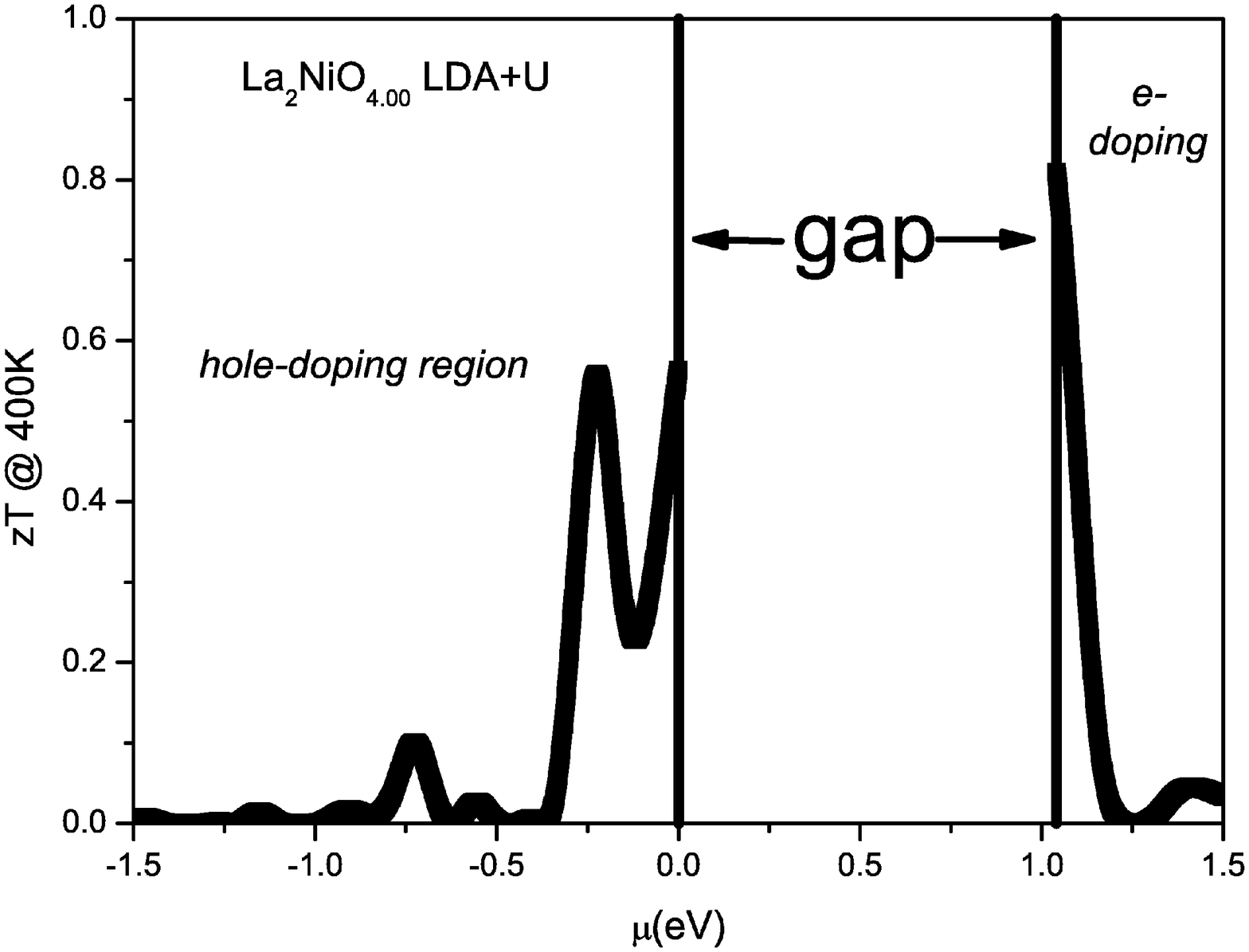}
\includegraphics[width=0.66\columnwidth,draft=false]{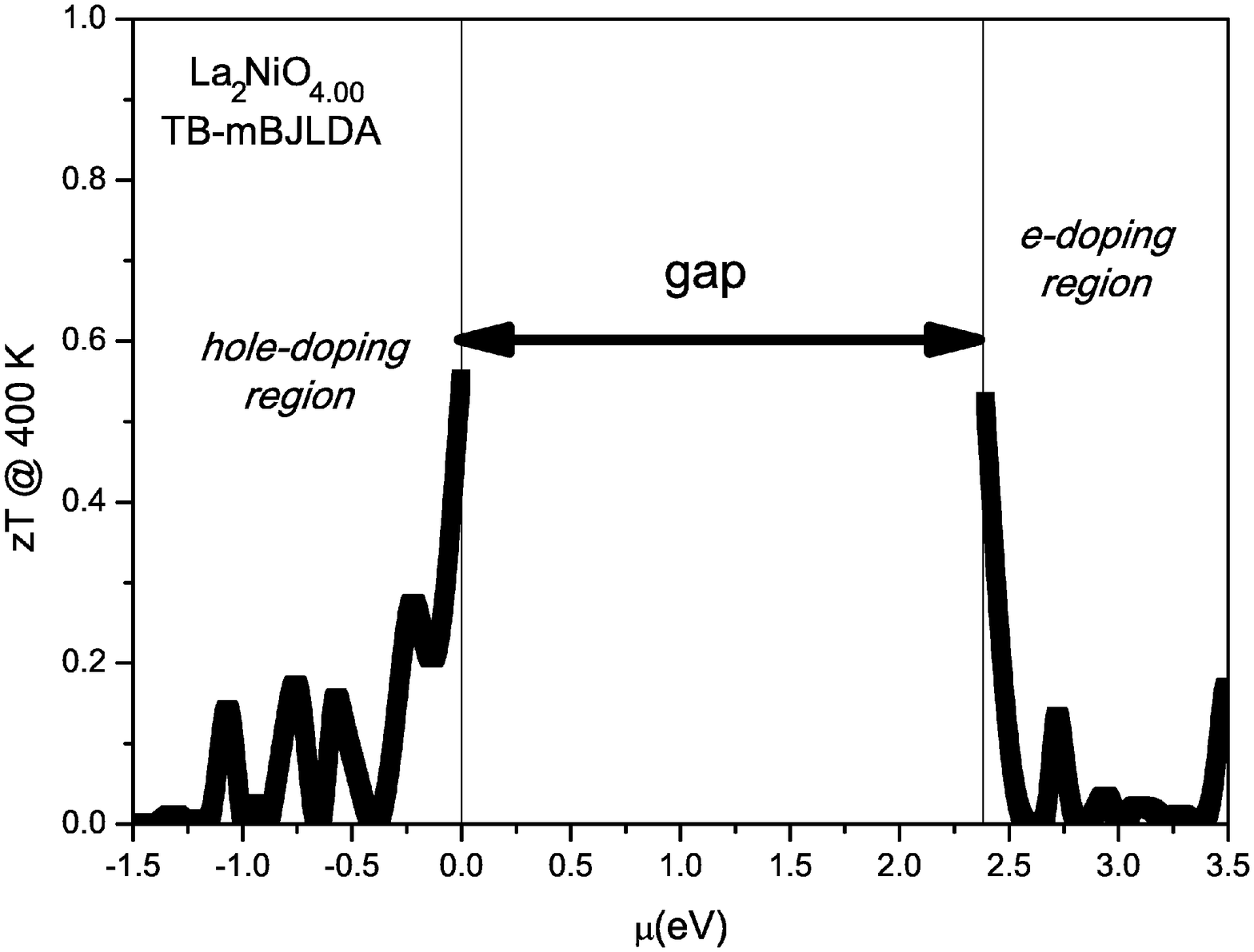}
\caption{Thermoelectric figure of merit (electronic part only) of La$_2$NiO$_4$ AF within GGA, the LDA+$U$ approach and TB-mBJLDA as a function of chemical potential ($\mu$= 0 represents the Fermi level). Observe a peak in the figure of merit for small hole-doping levels. $zT$ becomes zero at about -0.3 eV when the Seebeck coefficient changes sign from positive to negative, a similar feature happens in the electron-doped region, $S$ goes from negative to positive when the system is further electron doped.}\label{zT}
\end{center}
\end{figure*}

In order to see this in detail, we can start by analyzing the electronic transport properties of the parent compound. We have calculated these for La$_2$NiO$_4$ using various exchange-correlation potentials. Our calculations were carried out using the semiclassical approach, within the constant scattering time approximation. Yet, the value of the thermoelectric figure of merit $zT$= T$\sigma$S$^2$/$\kappa$ is independent of the scattering time chosen and so are the thermopower and $\sigma$/$\kappa$, as long as the constant scattering time approximation is retained. In Fig. \ref{zT} we present the thermoelectric figure of merit as a function of the chemical potential calculated at 400 K using different exchange-correlation functionals. We study the electronic part only, we do not consider phonon terms in the thermal conductivity, which might be substantial in particular for the localized electron limit. Thus, this electronic-only $zT$ can be considered as a theoretical upper limit of the figure of merit for this system. We will see below how we can make an estimate of these effects based on the experimental data for $\sigma$ and $\kappa$ from Refs. \onlinecite{la2nio4_sigma_singlecrystals}, \onlinecite{la2nio4_films_sigma} and \onlinecite{la2nio4_kappa}.

\begin{figure}[ht]
\begin{center}
\includegraphics[width=\columnwidth,draft=false]{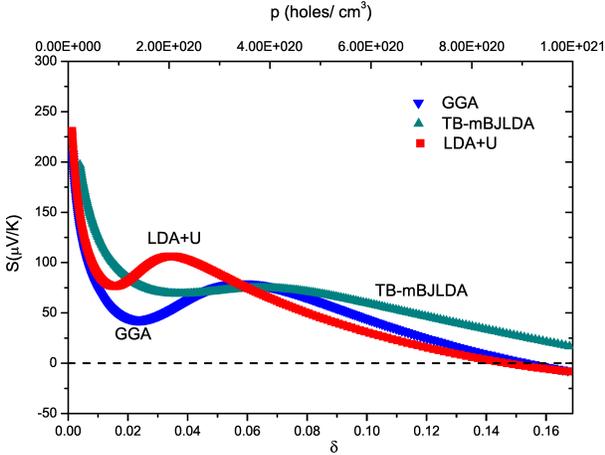}
\caption{(Color online.) Calculated thermopower at T= 400 K as a function of hole doping in La$_2$NiO$_{4+\delta}$ AF within GGA, the LDA+$U$ approach and TB-mBJLDA .}\label{S_vs_doping}
\end{center}
\end{figure}

Figure \ref{zT} shows a reasonably large value of the thermoelectric figure of merit for the stoichiometric compound (chemical potential at the top of the valence band), which is given by the large positive Seebeck coefficient, in qualitative agreement (see the discussion below) with experimental observations at low-$\delta$ (oxygen excess) values.\cite{la2nio4_Seebeck_delta} However, this is expected to be hampered by the large thermal conductivity due to phonons ($\kappa_{ph}$), which would substantially surpass its electronic counterpart ($\kappa_{el}$), which is the only component we are analyzing with these calculations. In any case, moving towards the itinerant electron limit, when $\kappa_{el}$ starts being comparable to $\kappa_{ph}$, these results become more significant. We observe an interesting peak in the figure of merit in the hole-doping region (oxygen excess, positive values of $\delta$, negative chemical potential). This peak is present independently of the exchange-correlation functional used, but the characteristics (height and position in chemical potential) vary depending on the method used.

In this paper we try to describe a physical phenomenon that occurs in a range of physical parameters (in this case $\delta$, the oxygen excess) in La$_2$NiO$_{4+\delta}$ but the quantitative values of which are very difficult to obtain ab initio. The peak is related to the $x^2-y^2$ bands, as we can see comparing Fig. \ref{zT} with the DOS plots in Fig. \ref{dos} (the energy window of the large peak in $zT$ lies inside that band). We also observe that other peaks in $zT$ appear at higher hole-doping concentrations (more negative chemical potential), related in that case with the $d_{z^2}$ and t$_{2g}$ bands. Even though the $d_{z^2}$ band is certainly narrower, it does not provide a sufficiently high $zT$ compared to $x^2-y^2$ (a multipeak structure is predicted by the TB-mBJLDA functional) because it is merged in the same energy region as the t$_{2g}$ bands. If somehow band engineering could be done to bring the d$_{z^2}$ band closer to the Fermi level, the possible thermopower obtained could be enhanced taking benefit of the localized nature of the band, and at a significantly smaller doping level. The realization of this possibility looks difficult. It can also be oberved 
how $zT$ becomes zero at about -0.3 eV simultaneously with the Seebeck coefficient sign change from positive at low doping to negative at higher doping levels. A similar feature happens in the electron-doped region, $S$ goes from negative to positive when the system is further electron doped (about 0.2 eV above the bottom of the conduction band).

In order to characterize in more detail that first peak in $zT$ at $\delta$ $\neq$ 0, and see how the amount of oxygen excess (hole-doping) relates to a particular change in chemical potential in the system we present in Fig. \ref{S_vs_doping} the calculated thermopower of the hole-doped system at 400 K for all the functionals used. Since the behavior of $\sigma$/$\kappa$ is relatively monotonous both with temperature and doping, the behavior of the thermopower (its peaks and sign changes) gives the broad picture of how the figure of merit itself evolves. The dependence of the Seebeck coefficient with both the doping level and the $\delta$ value is shown. A peak can be observed in all the curves although it is more pronounced for LDA+$U$ and becomes broadened within TB\-mBJLDA. The height of the peak is related to the width of the $x^2-y^2$ band, which is narrowed by the LDA+$U$ method, and broadened by the TB-mBJLDA scheme. An intermediate situation is given by uncorrelated GGA. The doping level necessary to reach it depends on the relative position of that band with respect to the Fermi level in the calculations, GGA placing it further away than the other methods (see Fig. \ref{dos}, top panel). It can be seen that for LDA+$U$ the peak is located at $\delta$$\sim$ 0.03, whereas for GGA and TB-mBJLDA it appears at higher doping levels $\sim$ 0.06 and 0.07, respectively. In addition, the above described sign change of the Seebeck coefficient in the hole-doped region can be seen for both GGA and LDA+$U$ at $\delta$ close to 0.15 (larger, beyond the values of doping we are studying here, for TB-mBJLDA).  This has been observed experimentally\cite{la2nio4_Seebeck_delta} for values around $\delta$ $\sim$ 0.20, in good agreement with the result obtained from our calculations. All these calculations were obtained for the parent undoped compound La$_2$NiO$_4$ as a function of the chemical potential, what could be called a rigid-band approach. We will argue below how the picture is modified when VCA is applied.

\subsection{VCA calculations}

\begin{figure}[ht]
\begin{center}
\includegraphics[width=0.49\columnwidth,draft=false]{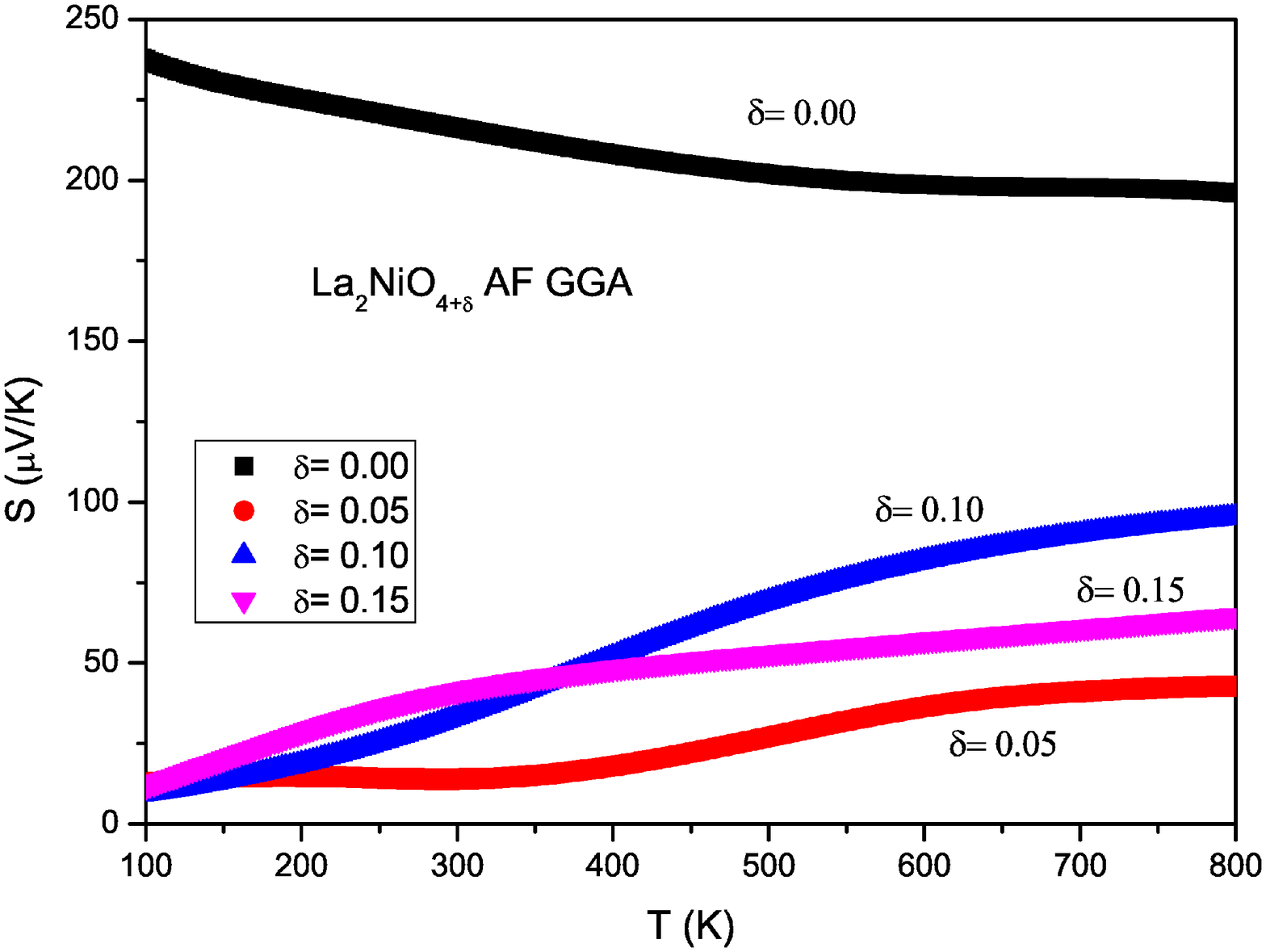}
\includegraphics[width=0.49\columnwidth,draft=false]{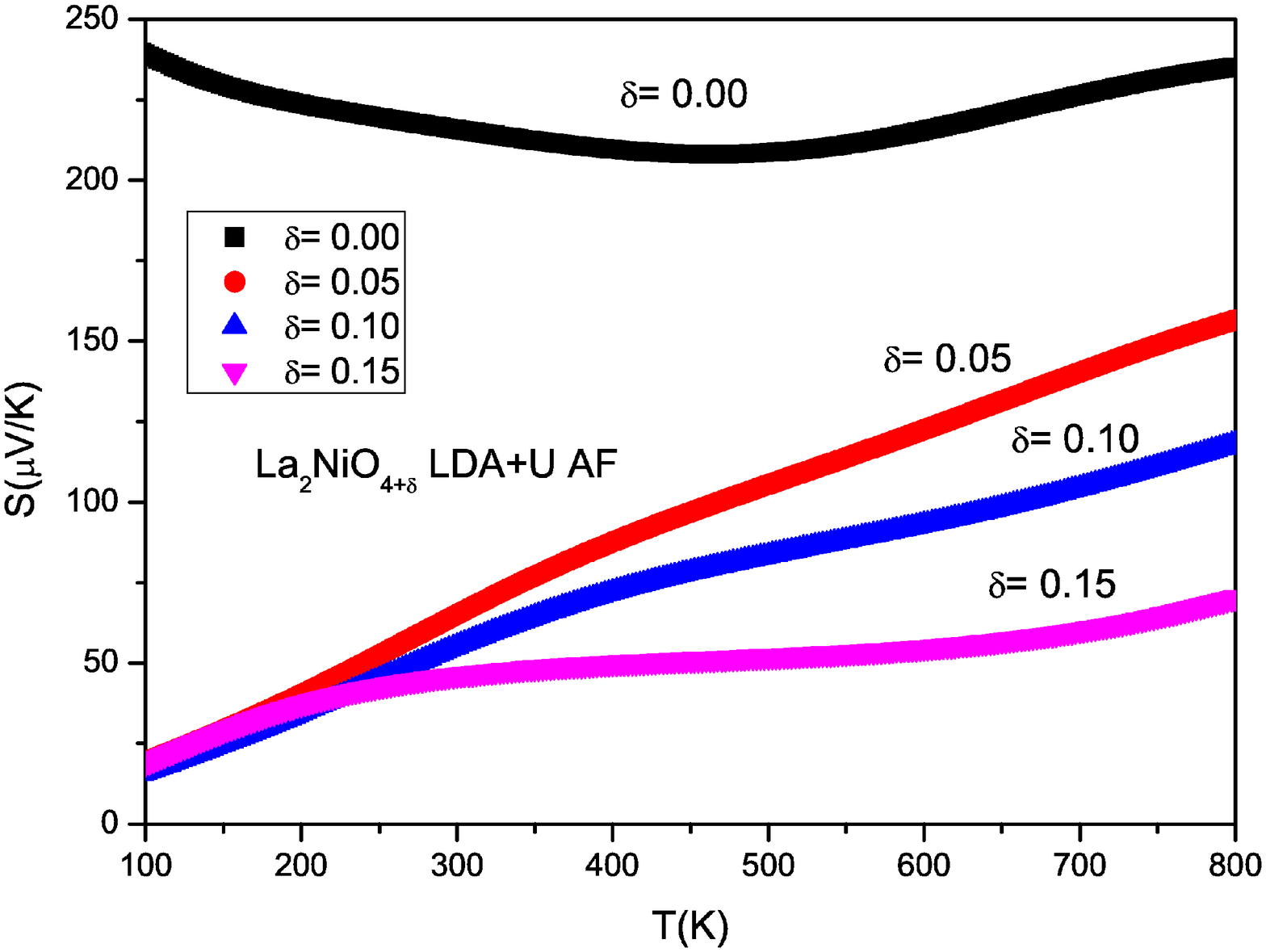}
\includegraphics[width=0.49\columnwidth,draft=false]{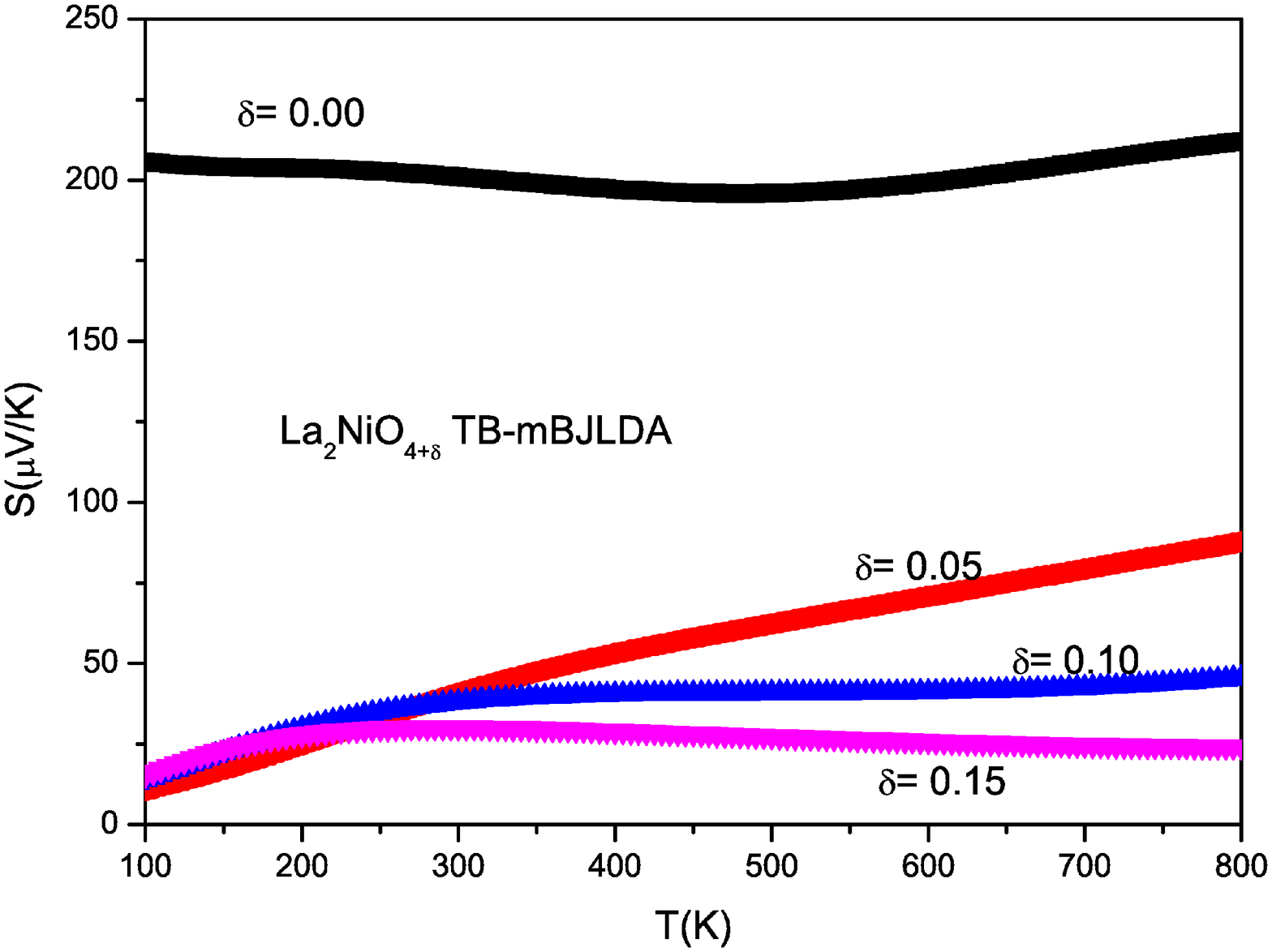}
\includegraphics[width=0.49\columnwidth,draft=false]{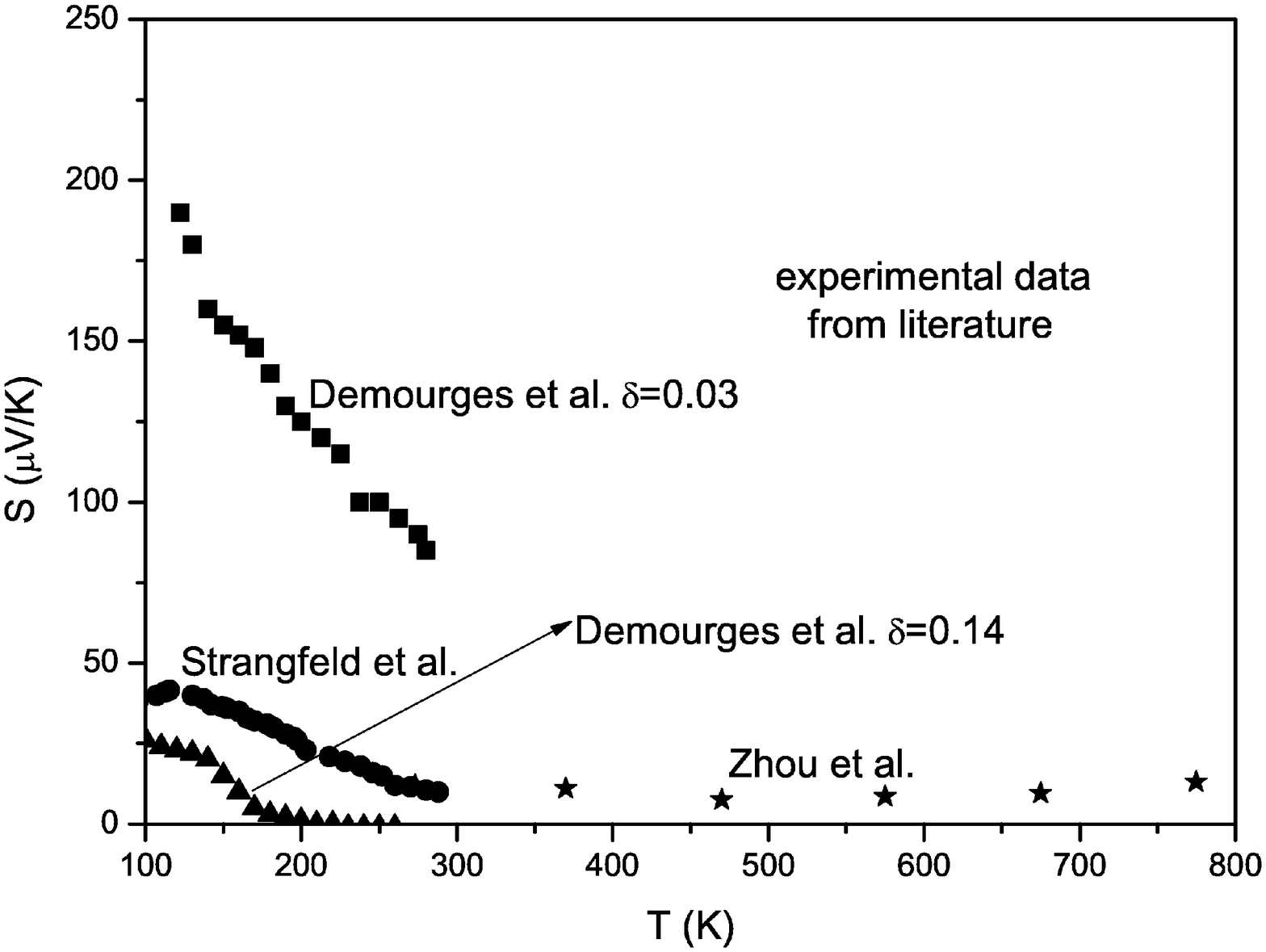}
\caption{(Color online). Thermopower of La$_2$NiO$_4$ AF within GGA, the LDA+$U$ approach and TB-mBJLDA and experimental values as a function of temperature for various values of the hole doping concentration $\delta$ simulated using the virtual crystal approximation. Observe the different tendency with temperature for optimum-doping values (increasing at high temperature) and higher values of doping (oscillating with temperature and flat at the high temperature end). These tendencies are consistent with experiments (shown in the lower right panel).}\label{S}
\end{center}
\end{figure}

\begin{figure*}[ht]
\begin{center}
\includegraphics[width=0.66\columnwidth,draft=false]{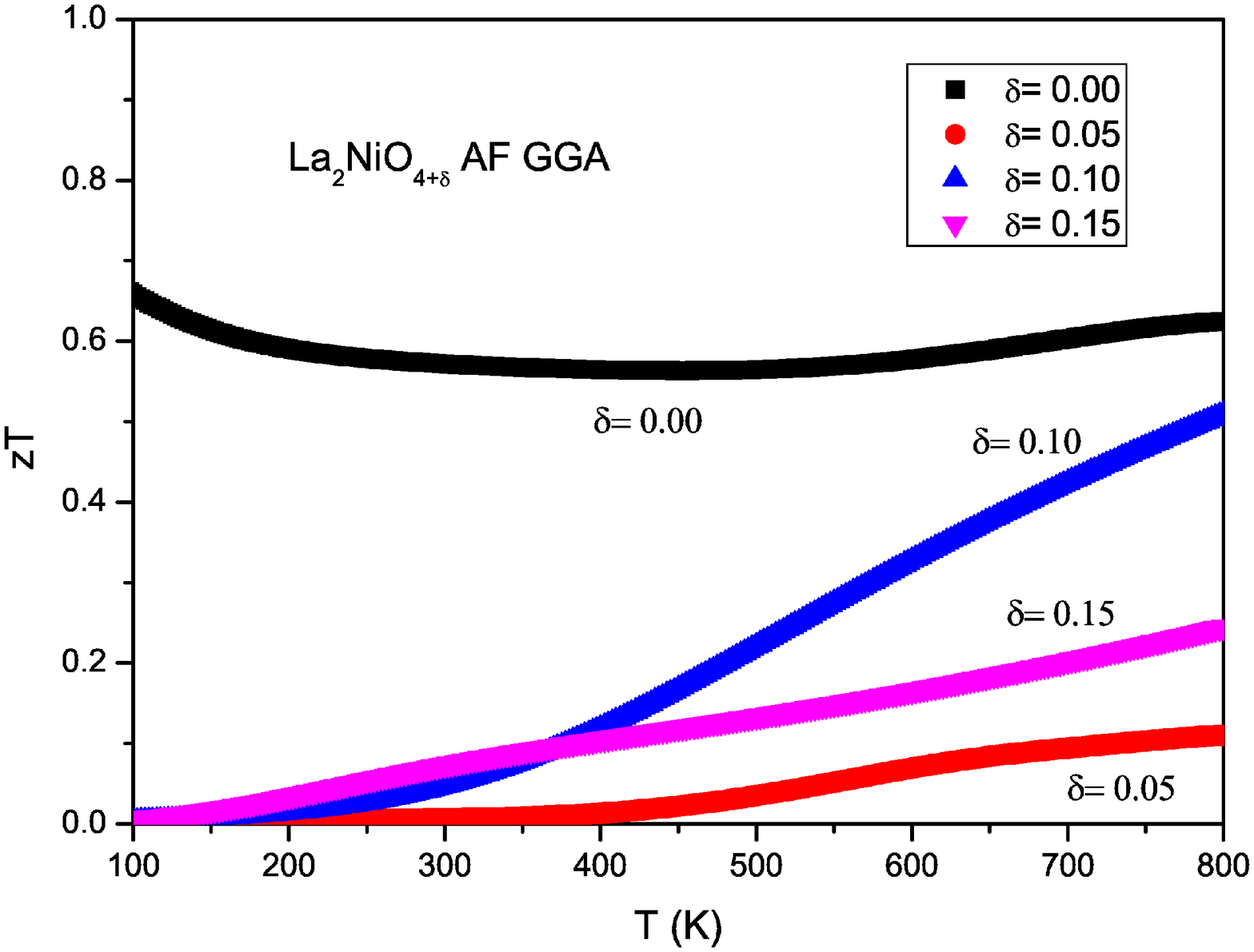}
\includegraphics[width=0.66\columnwidth,draft=false]{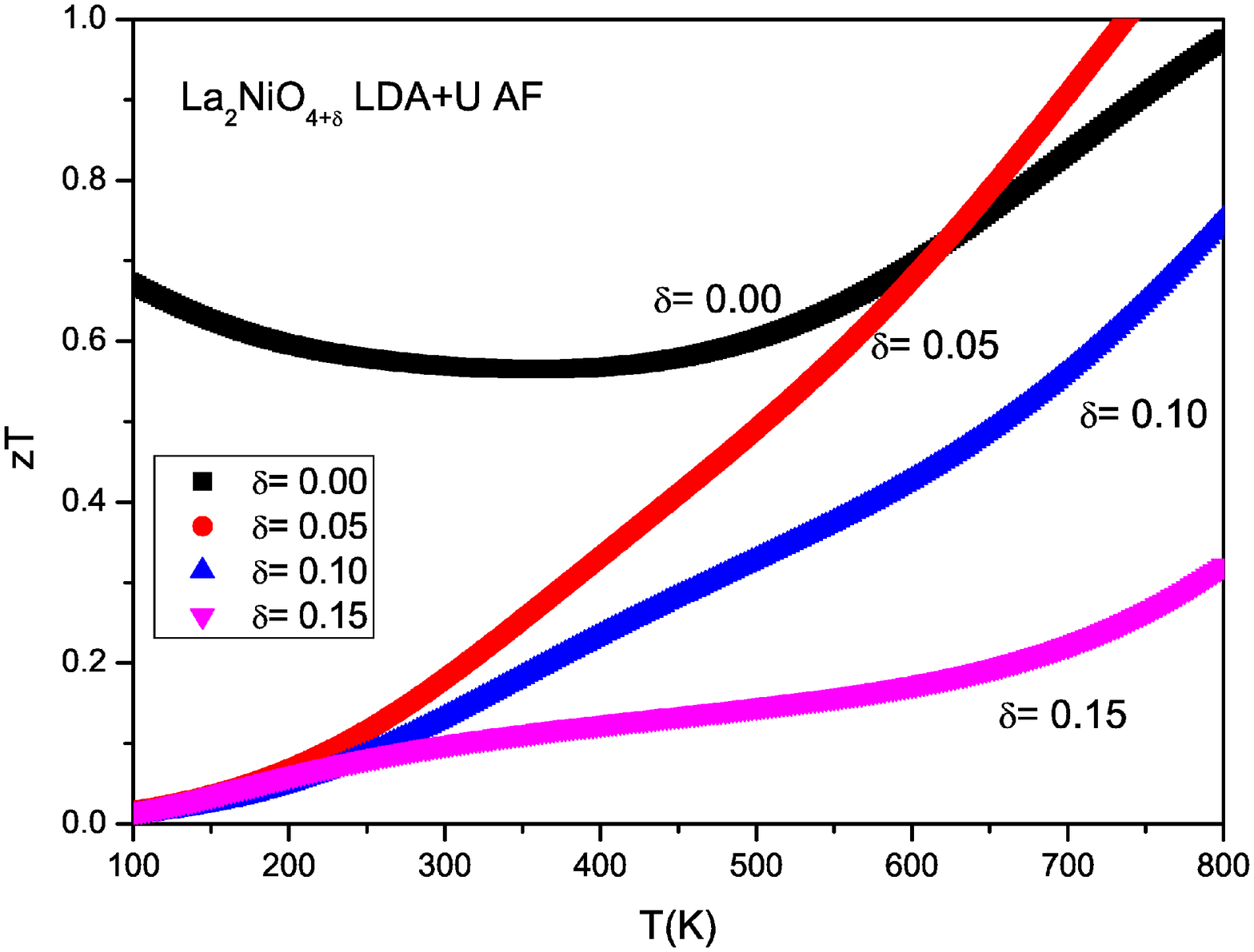}
\includegraphics[width=0.66\columnwidth,draft=false]{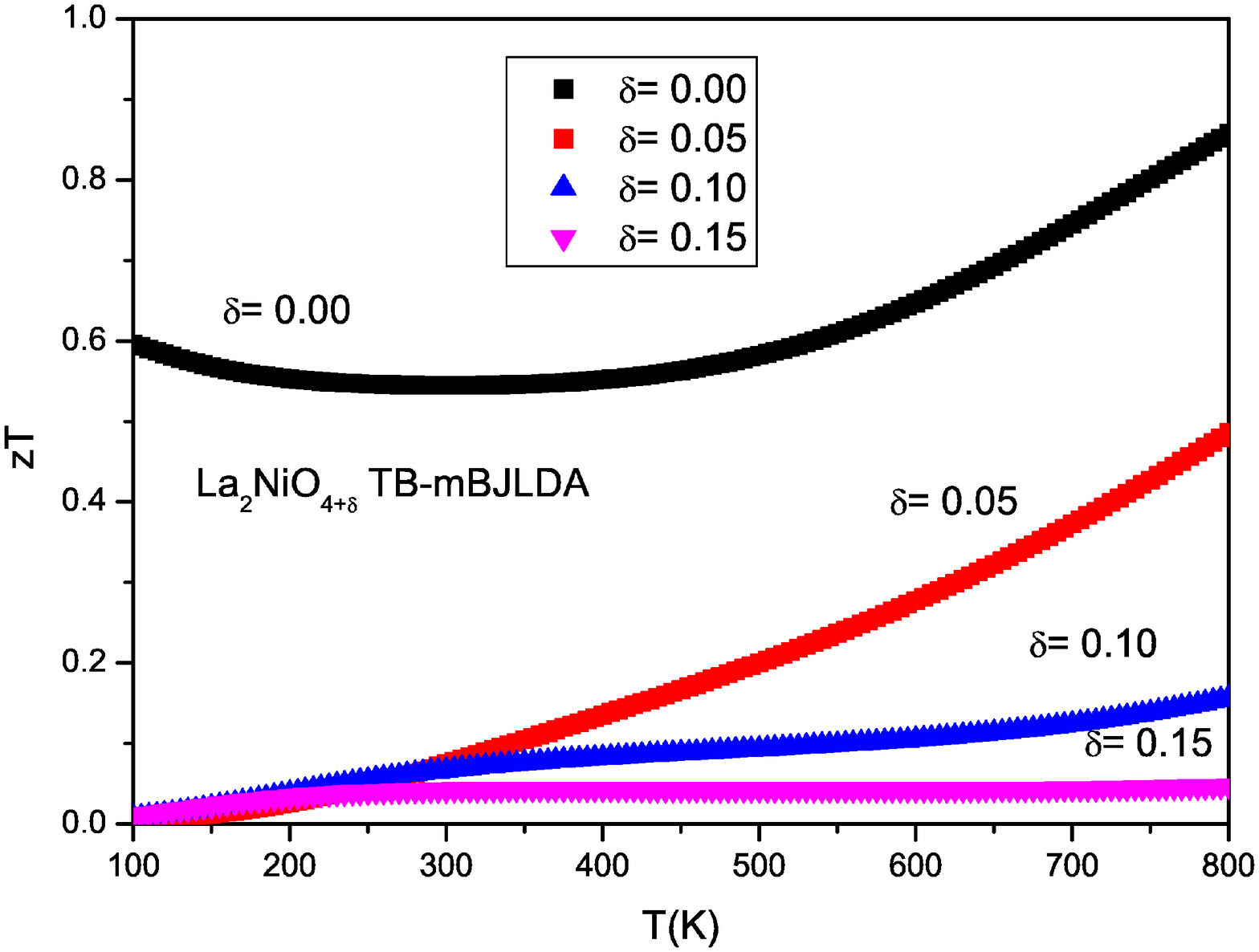}
\caption{(Color online). Thermoelectric figure of merit (electronic only) of La$_2$NiO$_4$ AF within GGA, the LDA+$U$ approach and TB-mBJLDA as a function of temperature for various values of the hole doping concentration $\delta$ calculated using the virtual crystal approximation. Observe the large high temperature figure of merit attainable at small doping values.}\label{zT_delta}
\end{center}
\end{figure*}

Once we have localized the position of the peak as a function of doping, we will study the evolution with hole-doping of the Seebeck coefficient (comparing it with available experimental data) and the other transport properties that can be calculated. In order to do that, we have performed VCA calculations (modifying the total number of electrons in the system to simulate the dopant concentration). Since we will be dealing with small doping levels ($\delta$ $\leq$ 0.15), we will consider the ground state AF configuration and also the structural properties of the parent compound. With this, we neglect the effects of the known orthorhombic distortion\cite{la2nio4_ortho1,la2nio4_ortho2} that occurs in this nickelate and would be a second order effect for the properties we try to estimate. The shape of the curves in Fig. \ref{zT} will not remain the same when VCA is applied because the effect of doping is not just a rigid band shift, but the existence of an optimum $zT$ with doping remains true. Also, the more pronounced maximum in figure of merit at negative values of the chemical potential occurs for the LDA+$U$ calculations, as we have seen in Figs. \ref{zT} and \ref{S_vs_doping} and will further confirm below. 


Figure \ref{S} shows the thermopower at different doping levels calculated within the various exchange-correlation potentials we are using, and an additional figure with some experimental data from the literature. We can observe large values of the Seebeck coefficient for the most insulating samples, where the thermopower is above 200 $\mu$V/K for the whole temperature range. The large Seebeck coefficient at low doping,\cite{la2nio4_Seebeck_delta} is well predicted by all functionals. In every case, the maximum in $zT$ occurs in the doping region of our interest ($\delta$ $\leq$ 0.15) for values that coincide roughly with the maxima in thermopower vs. $\delta$ at 400 K we analyzed in Fig. \ref{S_vs_doping} (below $\delta$ $\sim$ 0.05 within LDA+$U$ and at slightly higher values for GGA and TB-mBJLDA). For these ``optimum" doping values the thermopower increases at high temperatures while for the other $\delta$ values it shows an oscillating behavior and a flattening at higher temperatures, results that are compatible with most experiments available. The oscillating behavior of the thermopower with temperature that has been found in literature is particularly well described by a GGA calculation at $x$ $\sim$ 0.05 and also TB-mBJLDA at $\delta$ $\sim$ 0.10. According to our calculations, experiments showing thermopower oscillating with temperature (showing a maximum at intermediate temperatures) have been obtained analyzing samples in that doping region (hole-doped, non-stoichiometric, even if some of them do not explicitly mention the oxygen content).\cite{la2nio4_S_1,la2nio4_S_2,la2nio4_S_3} However, an accurate placement of the maximum in the Seebeck coefficient as a function of temperature is not given by our calculations, that type of agreement being similar to what has been found in other correlated oxides.\cite{sr6co5o15_antia} Zhou et al.\cite{la2nio4_S_1} have shown a slight increase in $S$ at high temperature experimentally, but the exact $\delta$ value of their samples is not provided; from our calculations we can argue that those samples would have a substantial $\delta$-value on the order of $\delta$ $\sim$ 0.15-0.2. The best quantitative agreement between our calculations and those reported from experiments at low-doping levels can be obtained with the TB-mBJLDA functional. This was also found for the thermoelectric properties of CrN, when compared with experiments and with LDA+$U$ calculations.\cite{crn_antia} This functional provides a purely ab initio parameter-free description of the system, not depending on the choice of an arbitrary parameter.

\begin{figure*}[ht]
\begin{center}
 \includegraphics[width=0.66\columnwidth,draft=false]{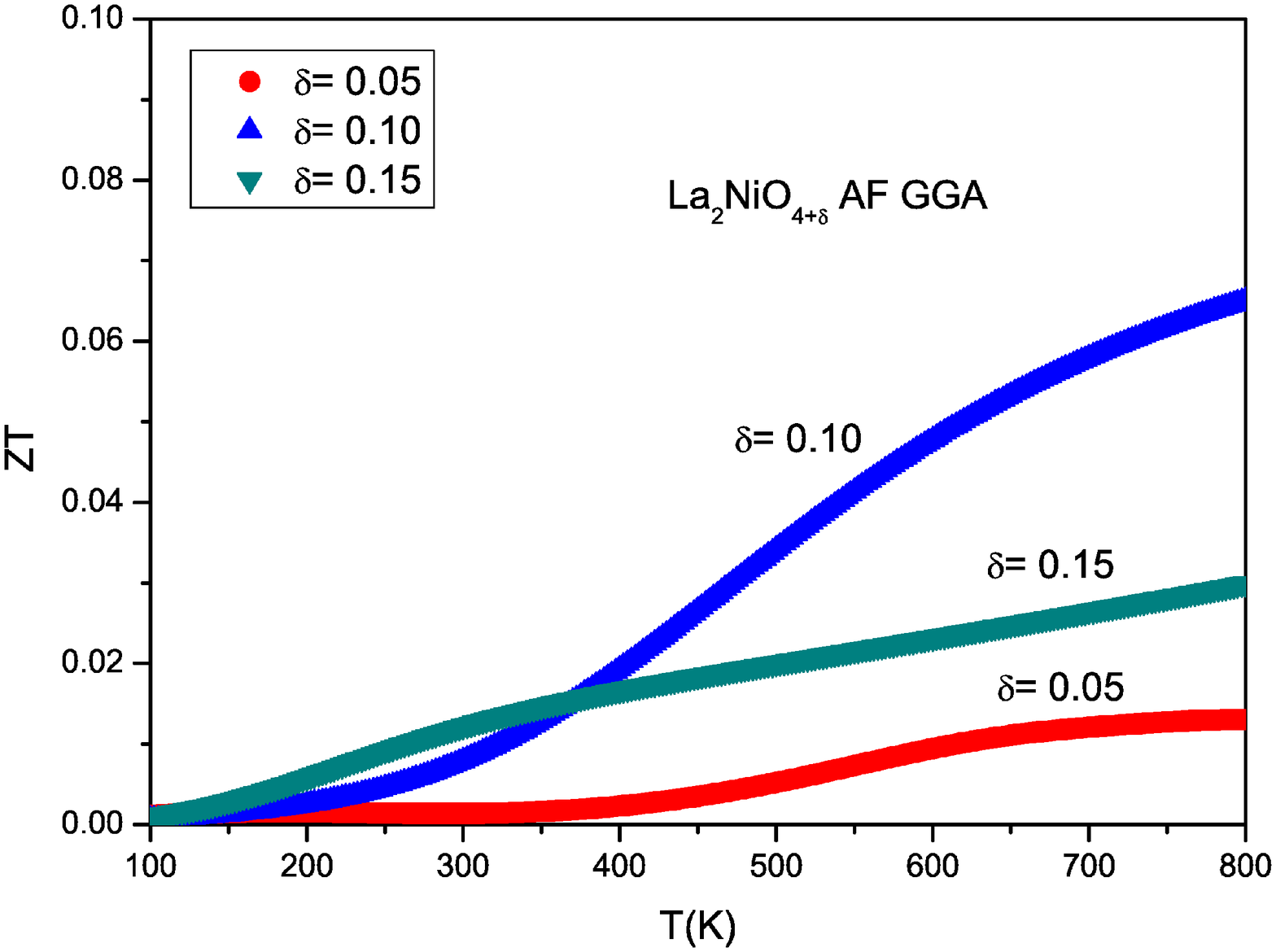}
\includegraphics[width=0.66\columnwidth,draft=false]{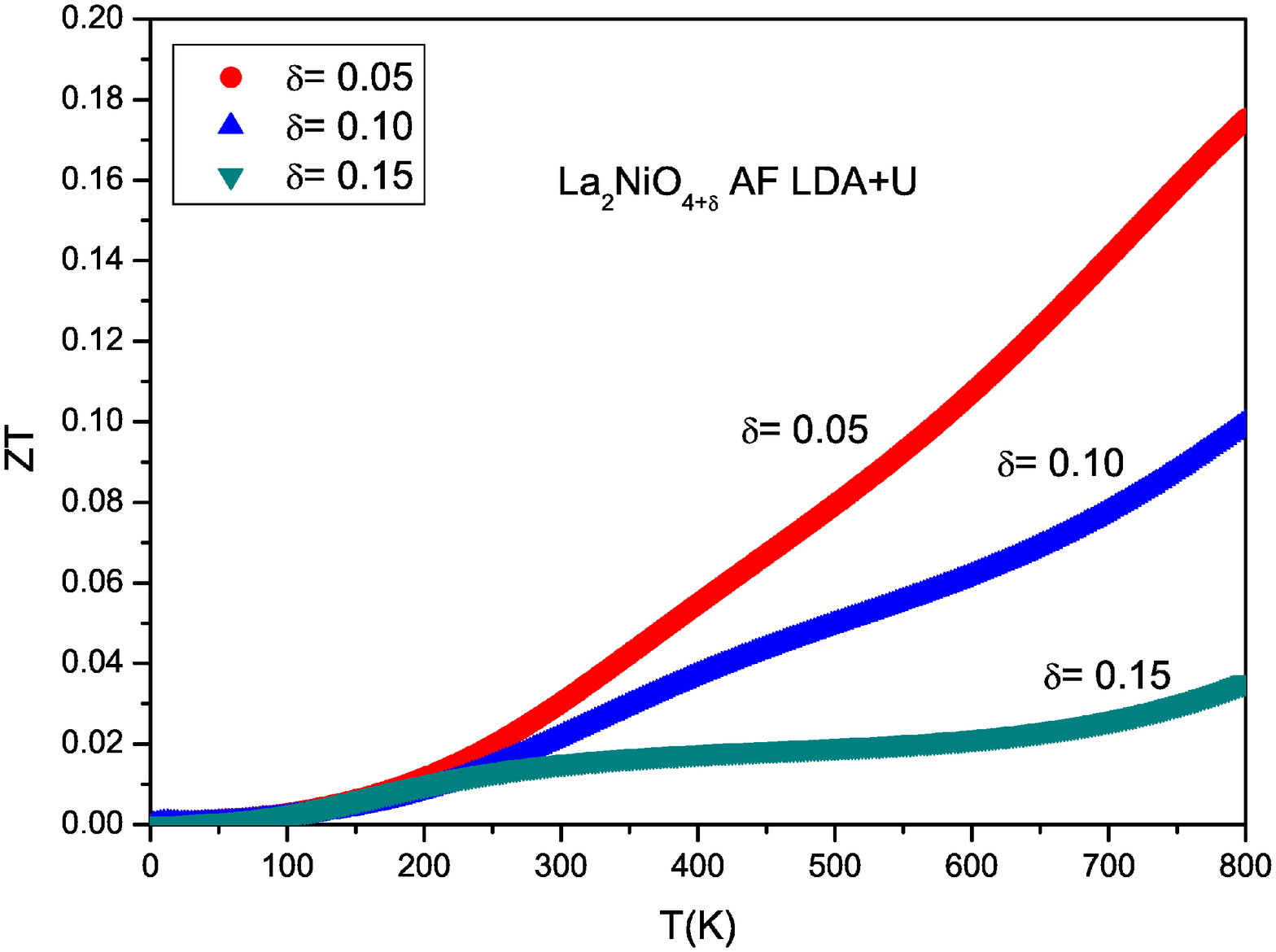}
\includegraphics[width=0.66\columnwidth,draft=false]{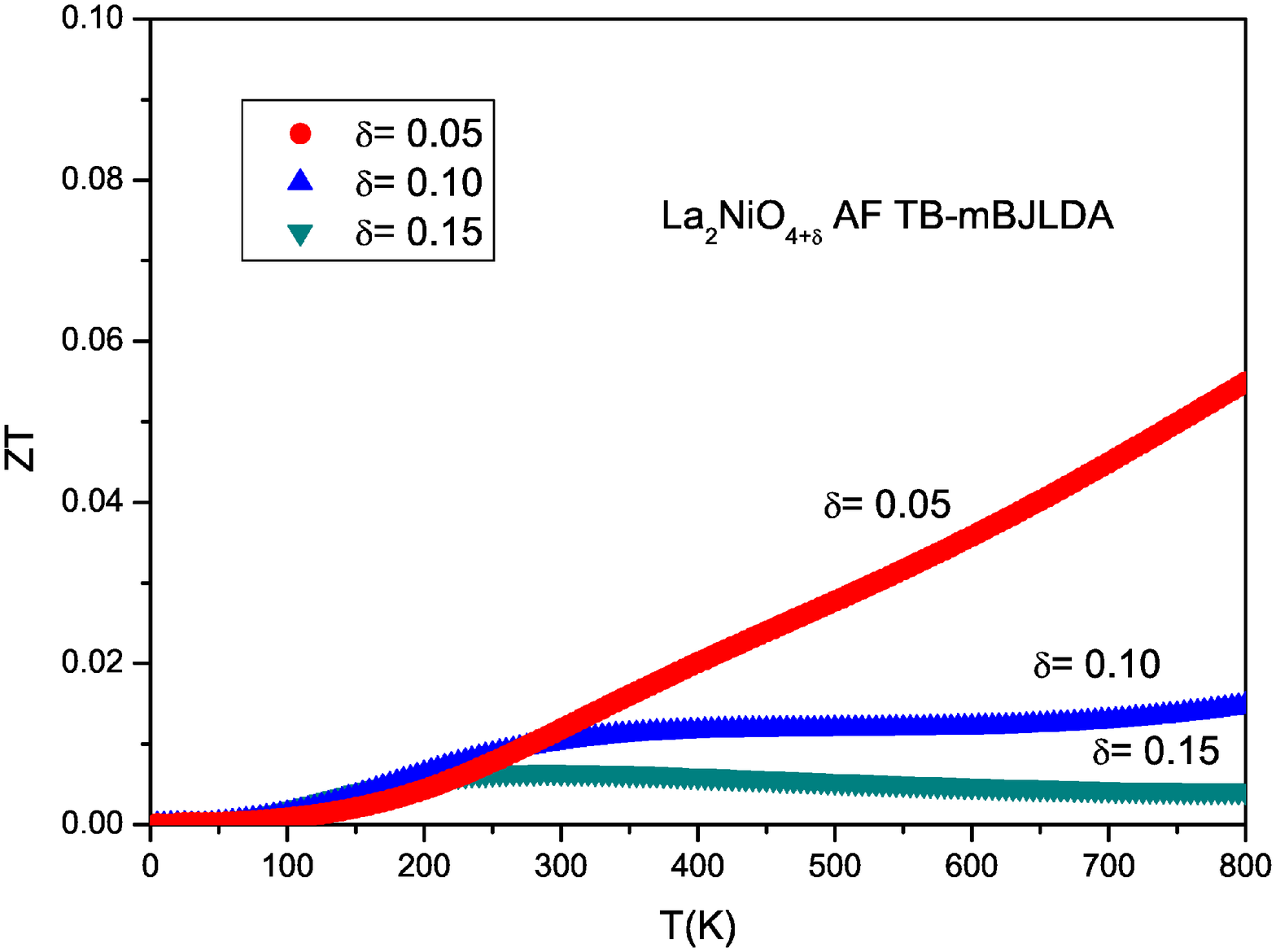}
\caption{Thermoelectric figure of merit  of La$_2$NiO$_4$ AF as a function of temperature for various values of the hole doping concentration $\delta$ within GGA, the LDA+$U$ approach and TB-mBJLDA estimated using the experimental values of $\kappa$ and $\sigma$}\label{zT_estimated}
\end{center}
\end{figure*}

In Fig. \ref{zT_delta} we present the evolution with temperature of the thermoelectric figure of merit (electronic-only) for various doping levels. The figure complements the description given by the thermopower analysis alone, with similar conclusions. LDA+$U$ values for $\delta$= 0.05 would be very promising in terms of the figure of merit obtained, even surpassing that of the undoped compound at high T. Also the values predicted by the parameter-free TB-mBJLDA functional would exceed $zT$= 0.5 at high temperatures for $\delta$= 0.05. Uncorrelated GGA predicts an optimum hole-doping content at $\delta$= 0.10 with large figure of merit at high temperature. LDA+$U$ also gives a very good response for $\delta$= 0.10 and 0.15, especially due to its high temperature dependence. From Figs. \ref{zT} and  \ref{S_vs_doping}, focused at 400 K and carried out with a rigid-band approach, we expected a sharper behavior in $\delta$-values, but the VCA results show a broader range of efficient thermoelectric response being predicted. We have already discussed that the large $zT$ of the $\delta$= 0 case would need to be reduced by introducing a substantial $\kappa_{ph}$. This would be particularly important at $\delta$= 0. Being the system an insulator, $\kappa_{ph}$ should be much higher than its electronic counterpart. Our calculations at $\delta$ $\neq$ 0 should be more realistic, being the system approaching the metallic limit, specially if the thermal conductivity due to phonons can be further reduced by nanostructuring. This has been shown in the past to be a key ingredient in the design of high performance thermoelectric devices.\cite{dresselhaus} In that case, $\kappa_{el}$ could become more important compared to $\kappa_{ph}$, as discussed above. Let us give some rough numbers. According to our calculations (within GGA, not shown), the conductivity increases by a factor 20 at room temperature for $\delta$= 0.05 with respect to the stoichiometric compound, and by a factor 80 for $\delta$= 0.15, explainable by the increase in carrier concentration that is also given in Fig. \ref{zT}. Also, if one can make thin films of the appropriate oxygen excess concentration, the conductivity can be made even larger, and the phonon thermal conductivity can be killed off by cutting away the long-wavelength phonon paths along the c-axis.

\subsection{More realistic estimates}

To further validate our results, we can give an estimate of the overall $zT$ (up to now, we have studied the electronic part only). Considering the Wiedemann-Franz relation, the figure of merit can be rewritten as: $zT$= $\sigma$TS$^2$/$\kappa$= $S^2$/$L_0$ being $L_0$ the Lorenz number with a value for free electrons $L_0= 2.45\times10^{-8} (V/K)^2$. At high enough temperatures, the lattice thermal conductivity term typically decreases as $1/T$ (with the electronic term roughly temperature independent). This type of behavior can be observed in Ref. \onlinecite{la2nio4_kappa}. Taking the values at the transition temperature for the thermal conductivity from Ref. \onlinecite{la2nio4_kappa}, the electrical conductivity from Ref. \onlinecite{la2nio4_films_sigma}, and using our calculated values for the thermopower, we can obtain a more realistic estimation of the $zT$ values for each functional at each doping level. The behavior of the estimated $zT$ with temperature for the various doping levels and functionals used can be seen in Fig. \ref{zT_estimated}. It can be observed that the shape of the curves for each functional remains roughly the same than in Fig. \ref{zT_delta}. The calculated electronic-only factor $\sigma$T/$\kappa$ is not far from constant when the real band structure of the material is introduced in the calculations. The overall values of $zT$ are reduced with respect to those obtained only with the electronic contribution in approximately one order of magnitude due to the phonon thermal conductivity. $z$T $\sim$ 0.1 is predicted from our calculations at high temperature if the proper oxygen content is chosen. In any case, the growth in the form of thin films with the appropriate oxygen content would increase the electrical conductivity and reduce the phonon thermal conductivity. Our results show that, in that case, the system could become promising for further research in terms of its thermoelectric properties, trying to find other oxides with performance comparable to that of layered cobaltates.

\section{Summary}\label{summary}

Our ab initio calculations for the compound indicate some promising features of hole-doped La$_2$NiO$_4$ as a possible oxide thermoelectric compound. If the thermal conductivity can be reduced by nanostructuring, e.g. in the form of thin films, the system could show enhanced thermoelectric performance at low hole-doping levels, attainable by the appropriate control of the apical oxygen content. Our calculations show that a region with relatively large Seebeck coefficient exists in this compound at small doping levels, within the realistic AF description. A careful experimental study needs to be performed in this respect controlling the oxygen content, and also making thin films with the appropriate oxygen composition. Hole-doping will increase the conductivity, as the thin film geometry does, which together with the reduction in thermal conductivity could leave room for an improvement of the thermoelectric figure of merit of this layered nickelate. In terms of the electronic structure, band engineering could be used to explore the possibility of using the large thermopower expected from the $d_{z^2}$ band being closer to the Fermi level.

\section{Acknowledgments}

We thank F. Rivadulla and C.X. Quintela for fruitful discussions. V.P. and A.S.B. acknowledge the Spanish Government for support through the Ram\'on y Cajal program and FPU program, respectively. Financial support was given from the Ministry of Science of Spain through project MAT-200908165.


\end{document}